\documentclass[aps,prl,twocolumn,showpacs,superscriptaddress,amssymb,amsfonts]{revtex4}

\usepackage{graphicx} 

\begin{document}
    
\title{Direct Measurement of the Photon Statistics of a Triggered Single Photon Source}
\author{F. Treussart}
\email{treussar@physique.ens-cachan.fr}
\author{R. All\'eaume}
\author{V. Le Floc'h}
\affiliation{Laboratoire de Photonique Quantique et Mol\'eculaire,
ENS Cachan, 61 Avenue du Pr\'esident Wilson, 94235 Cachan Cedex, France}
\altaffiliation{Laboratoire du CNRS, UMR 8537, associ\'e à l'\'Ecole Normale 
Sup\'erieure de Cachan}
\author{L.T. Xiao}
\altaffiliation{Permanent address: Dept. of Electronics \& Information Technology, 
Shanxi University, Taiyuan, China}
\affiliation{Laboratoire de Photonique Quantique et Mol\'eculaire,
ENS Cachan, 61 Avenue du Pr\'esident Wilson, 94235 Cachan Cedex, France}
\author{J.-M. Courty}
\affiliation{Laboratoire Kastler Brossel, UPMC case 74, 4 place Jussieu, 
75252 Paris Cedex 05, France} 
\altaffiliation{Laboratoire de l'\'Ecole 
Normale Sup\'erieure et de l'Universit\'e Pierre et Marie Curie, 
associ\'e au CNRS.}
\author{J.-F. Roch}
\affiliation{Laboratoire de Photonique Quantique et Mol\'eculaire, 
ENS Cachan, 61 Avenue du Pr\'esident Wilson, 94235 Cachan Cedex, France}
\date{\today}

\begin{abstract}
We studied intensity fluctuations of a single photon source relying 
on the pulsed excitation of the fluorescence of a 
single molecule at room temperature. We directly
measured the Mandel parameter $Q(T)$ over 4 orders of magnitude of observation 
timescale $T$, by recording every photocount. 
On timescale of a few 
excitation periods, subpoissonian statistics is clearly observed 
and the probablility of two-photons events is 10 
times smaller than Poissonian pulses. 
On longer times, blinking in the fluorescence, due to the molecular triplet state, 
produces an excess of noise.
\end{abstract}

\pacs{42.50.Dv, 03.67.Dd, 33.50.-j}

\maketitle

Over the past few years, there has been a growing interest for 
generating a regular stream of single photons on demand. This was mainly motivated by 
applications in the field of quantum cryptography \cite{Gisin_RMP01}.
An ideal single photon source (SPS) should produce light  pulses
containing exactly one photon per pulse, triggered with a repetition 
period $\tau_{\rm rep}$, and delivered at the place of interest with 
100\% efficiency. For any given measurement time $T$, this source would 
emit exactly $N=T/\tau_{\rm rep}$ photons, so that the standard 
deviation $\Delta N\equiv \sqrt{\langle N^2\rangle_{T} - \langle N\rangle_{T}^2}=0$ ($\langle\;\rangle_{T}$ 
has to be understood as a mean value over a set of measurements lasting $T$).
Such a source would then be virtually free of intensity 
fluctuations, therefore corresponding to perfect intensity squeezing \cite{Reynaud_90}. 

A first category of SPSs already realized consists of sources operating 
at cryogenic temperature.
They rely on optically \cite{Kim_99,Michler_Science_00,Santori_PRL_01,Moreau_APL_01} or 
electrically \cite{Yuan_02} pumped semiconductor
nanostructures or on the fluorescence of a two level system coherently 
prepared in its excited state \cite{Brunel_PRL99}.
A one--atom micromaser has also been used to prepare arbitrary photon number states on 
demand \cite{Brattke_PRL01}. 
However the collection efficiency of photons is barely higher than a few 
$10^{-3}$ in these experiments. 
Due to this very strong attenuation, the intensity statistics are 
very close to a Poisson law at the place where the stream of photons is available.
Another route is to realize SPSs at room temperature. In this case higher collection 
efficiency (around 5\%) is achieved. The existing room--temperature 
SPSs rely on the pulse saturated emission of a single 4--levels 
emitter \cite{DeMartini96,Rosa_PRA00}. 

When the pulse duration $\tau_{\rm p}$ is much shorter than 
the dipole radiative lifetime $\tau_{\rm rad}$, such a single 
emitter can only emit one photon per pulse. This temporal control of the dipole excitation 
allows therefore to easily produce individual photons on demand \cite{Lounis00,Alexios_EPJD}.
However, in previous SPS realizations, little attention has been paid to 
analyse their intensity fluctuations. 
To address this problem we realized a room--temperature SPS 
relying on the pulsed saturation of a single molecule embedded in a thin 
polymer film \cite{FMT_OL01}. 

The samples are made of cyanine dye DiIC$_{18}$(3) molecules dispersed at a 
concentration of about one molecule per 10~$\mu m^2$ into a 30~nm thick PMMA layer, spincoated over a microscope coverplate.
The fluorescence from the single molecule is excited and collected by 
the standard technique of scanning confocal optical 
microscopy \cite{Nie_Zare97}.
The molecules are non--resonantly excited at 532~nm, with femtosecond 
pulses generated by a Ti:Sapphire laser and frequency 
doubled by single pass propagation into a LiIO$_{3}$ crystal. The repetition rate, initially at 82~MHz, is divided by a pulsepicker.
The energy per pulse $E_{\rm p}$ is adjustable by an electro--optic 
modulator. The pulse duration is about $\tau_{\rm p}\approx 100$~fs.
The excitation light is reflected by the dichroic mirror of an
inverted microscope, and then focused by an oil--immersion objective 
($\times 60$, NA=1.4), to form a spot of $\approx 1\mu$m$^2$ surface area.
The fluorescence light from the sample is collected by the same objective and then focused 
into a 30~$\mu$m diameter pinhole. 
After recollimation, a holographic notch filter 
removes the residual pump light. A standard Hanbury Brown and 
Twiss (HBT) setup is then used to split the beam and detect single photons on two identical avalanche 
photodiodes.
Glass filters are placed onto each arm to suppress parasitic crosstalk 
\cite{Kurtsiefer_01} between the two photodiodes.

\begin{figure}[h]
\includegraphics[width=8cm]{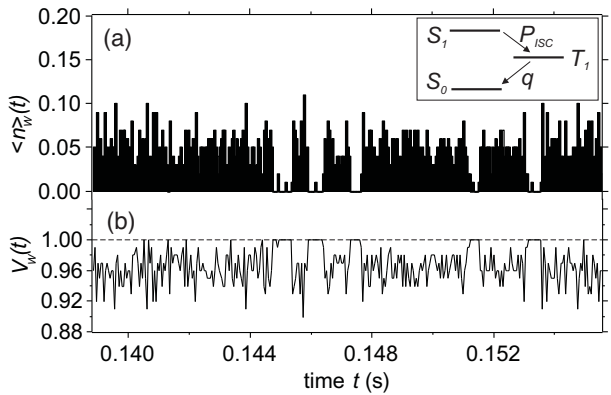} 
\caption{(a) Mean number $\langle n\rangle_{\rm w}$ 
of photons detected in successive samples of the data of constant size 
$W=100$ cycles (bin period of $50~\mu$s). 
(b) Corresponding time trace of the normalised variance $V_{\rm w}$ 
showing reduction of noise ($V_{\rm w}<1$) during the emission period of the molecule.
Inset of (a) displays the molecular level diagram.}
\label{variance_vs_t}
\end{figure}

In order to rapidly identify single molecule emission, we first measure the intensity 
autocorrelation function $\langle I(t)I(t+\tau)\rangle$ of the 
fluorescence light by the standard Start--Stop 
technique with a time--to--amplitude converter \cite{Brunel_PRL99}. 
When a single emitter is addressed, there is
virtually no event registered at $\tau=0$, since a single photon cannot be 
simultaneously detected on both sides of a 
beamsplitter \cite{Grangier_EPL86}.
The histogram shows a peak pattern at the pulse repetition period 
$\tau_{\rm rep}$.
As explained in Ref.\cite{Brunel_PRL99}, the peaks' areas allow one 
to infer the probabilities $P_{\rm S}(n)$ for the source (S), of giving 
$n=0,1,2$ photocounts per excitation pulse, where 2 photons counts 
are due to deviation from the ideal SPS. Nevertheless, this technique can hardly 
be used to extract the intensity fluctuations on timescale longer than a single pulse.
We have therefore chosen to record each photodetection event with
a two--channel Time Interval Analyser computer board (GuideTech, Model GT653).
Since each detection channel has a deadtime of 250~ns the 
excitation repetition rate was chosen to be 2~MHz.
\begin{figure}[h]
\includegraphics[width=8cm]{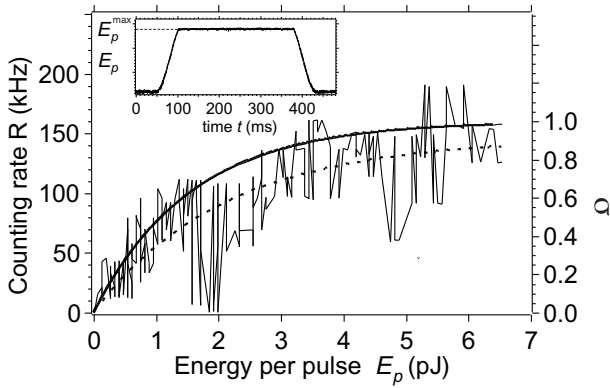} 
\caption{Photon counting rate $R$ vs. energy per pulse, for a single 
molecule. The inset shows the excitation ramp $E_{\rm p}(t)$, with 
$E_{\rm p}^{\rm max}=13$~pJ in this case.
The record of the saturation curve was limited by the photobleaching of the 
dye. The dashed curve is a fit of the raw data according to 
eq.(\ref{sat_TLS}) and the solid line is a fit after correction of triplet state 
excursion. The right scale shows the excited state population $\sigma$.} 
\label{saturation}
\end{figure}
In a typical experiment, we first raster scan the sample at low 
excitation energy per pulse (0.5~pJ). 
When a single molecule is located, we apply the excitation energy ramp 
$E_{\rm p}(t)$ shown on inset of Fig.\ref{saturation}, and simulteaneously record the 
fluorescence counts on a 1~ms integration time. Fig.\ref{saturation} displays 
the fluorescence counting rate $R$ vs. $E_{\rm p}$. The large intensity fluctuations are due to triplet state excursion 
of the molecule (see Fig.\ref{variance_vs_t}). If this state 
is not taken into account, the molecular energy levels can be 
modelised by a 2-level system assuming a very fast non-radiative relaxation between the two 
higher and the two lower energy states. The excited state population $\sigma$ at the time $\tau_{\rm 
p}$ after the pulse arrival is then
\begin{equation}
\sigma= {E_{\rm p}/E_{\rm sat}\over{(1+E_{\rm p}/E_{\rm sat})}}
\left( 1-e^{-{\tau_{\rm p}\over\tau_{\rm rad}}\left(1+{E_{\rm p}\over E_{\rm 
sat}}\right)} \right),
\label{sat_TLS}
\end{equation}
where $\tau_{\rm rad}\approx 2.8~$ns for the cyanine considered. 
The data $R(E_{\rm p})$ are fitted by the function $R=R_0\times \sigma$ 
in a two steps procedure. After a first fit of the 
raw data, all the points below this fit, which are attributed to triplet state 
excursion, are removed. The fit of the remaining set of data yields
$R_{0}=160\times 10^{3}$ counts/s and $E_{\rm 
sat}=5.6\times 10^{-5}$~pJ. 

In order to optimize the number of emitted photons and avoid rapid photobleaching, we 
then set $E_{\rm p}^{\rm max}$ to 5.6~pJ. Such a value would correspond to 
$\sigma=97\%$, for the molecule studied in Fig.\ref{saturation}.
During the constant maximum pumping energy period of the excitation 
ramp, $10^4$ detection events 
are typically recorded before photobleaching. Thanks to the high stability of the period of 
the pulsed laser, this set of times can be synchronized on an 
excitation timebase. We then build the table of the number of photocounts $n_i=0,1,2$ 
for each excitation pulse $i$.
Photons which are delayed by more than $10\times\tau_{\rm rad}$ 
are considered to come from the dark counts of the two photodiodes, and are 
therefore rejected. 
\begin{table}
\begin{tabular}{p{.7cm}p{1.5cm}p{2cm}p{1.5cm}p{1.5cm}}
X & $P_{\rm X}(1)$& $P_{\rm X}(2)$& $\overline{n}_{\rm X}$& $V-1$\\
\hline\hline
S & 0.0466& $5.0\times10^{-5}$& 0.0467& -0.0445\\
R & 0.0452& $50\times10^{-5}$& 0.0462& -0.0244 \\
C & 0.0451& $53\times10^{-5}$& 0.0462& -0.0231\footnote{calculated from $P_{\rm C}(n)$}\\
\hline\hline
\end{tabular}
\caption{Photocount probabilities $P_{\rm X}(n),\; n=1,2$, of our SPS (X=S), 
of an experimental reference source (R) and 
of a theoretical coherent source (C), the photocount statistics of which 
is affected by the detection setup. 
$\overline{n}_{\rm X}$ are the mean number of detected photons per pulse and 
$V$, the normalized variances. Negative values of $V$ for the coherent and 
reference sources are due to a dead time effect.}
\label{tab_compar}
\end{table}

The data considered hereafter corresponds to a molecular source (S) which
survived during 319769 periods (about 160~ms)
yielding 14928 recorded photons including 14896 single photon events, 16 two-photons 
events. We deduced $P_{\rm S}(1)=0.0466$ and $P_{\rm S}(2)=5.0\times 10^{-5}$ 
and a mean number of detected photon per pulse $\overline{n}_{\rm S}= 
0.0467$ (see Table \ref{tab_compar}).
The real source is considered as the 
superposition of an attenuated ideal SPS with an overall quantum efficiency 
$\eta$, and a coherent source simulating the background,
which adds a mean number of detected photon per pulse $\gamma$. From the measured 
values of $P_{\rm S}(1)$ and $P_{\rm S}(2)$, we infer $\eta\approx 
0.0445$ and $\gamma\approx 2.2\times 10^{-3}$. This leads to a signal-to-background ratio of about 
20.

We also compared experimentally our SPS to a reference
source (R) made of attenuated pump laser pulses, with
approximately the same mean number of detected photons per pulse. 
Quantitative tests of this reference source and of the detection setup are however necessary. 
A particular care has to be taken to the bias of photocount statistics 
due to the detection deadtimes on both channels.
For a coherent state of light (C) containing $\alpha$ photon per 
excitation pulse, one can readily calculate the counting probability distribution
and show that $\displaystyle P_{\rm C}(0)=e^{-\alpha}, P_{\rm C}(1)=2e^{-\alpha/2}(1-e^{-\alpha/2})$, 
$\displaystyle P_{\rm C}(2)=(1-e^{-\alpha/2})^{2}$ and $\overline{n}_{\rm C}=2(1-e^{-\alpha/2})$, 
where $\overline{n}_{\rm C}$ is the mean number of detected photons per pulse.
For the reference source (R), we measured $\overline{n}_{\rm R}=0.0462$, 
$P_{\rm R}(1)=0.0452$, $P_{\rm R}(2)=50\times 10^{-5}$, whereas one 
predicts, for $\overline{n}_{\rm C}=\overline{n}_{\rm R}$, $P_{\rm C}(1)=0.0451$ and 
$P_{\rm C}(2)=53\times 10^{-5}$. The measured values are in good agreement with the predictions, which 
proves that the faint Ti:Sapphire pulses make a good calibration source for Poisson statistics.
We then infer the ratio $P_{\rm S}(2)/P_{\rm R}(2)=0.10$, which tells 
that the number of two photons pulses in our SPS, is 10 times smaller  
than in the reference poissonian source (R).

In a first attempt to estimate the fluorescence intensity fluctuations per 
pulse, we considered samples of the data made of $W$ excitation cycles.
We introduced a normalized variance $V_{\rm W}$ defined, on the sample, by
$V_{W}\equiv \langle(\Delta n)^{2}\rangle_{\rm w}/\langle n\rangle_{\rm w}$, with
$\langle(\Delta n)^{2}\rangle_{\rm w}\equiv\sum_{i=1}^{W} (n_i-\langle 
n\rangle_{\rm w})^{2}/W$, where $n_i$ is the number of detected photons for the pulse $i$ 
and $\langle n\rangle_{\rm w}$ is the mean number of detected photon per pulse in 
the sample considered. In the very few samples for which $\langle n\rangle_{\rm w}=0$, $V_{W}$ is 
not defined and is set to 1. For a Poisson distribution of photocounts $V_{W}=1$,
whereas $V_{W}<1$ for subpoissonian distribution. 

In order to follow the time evolution of intensity fluctuations, 
we then extract from the set of $\lbrace n_i\rbrace$ all the successive samples of photocounts measurements of size $W=100$,
separated by a single pulse period.
Fig.\ref{variance_vs_t}(a) displays the mean number of detected photons 
per pulse vs. time. We clearly see random intermittency in the fluorescence of the 
molecule, due to the presence of a dark triplet state $T_{1}$ in the molecular 
energy levels diagram (see inset of Fig.\ref{variance_vs_t}(a)). 
At each excitation cycle the molecule has a small probability 
to jump into this non--fluorescent state, where it stays for a time 
much longer than the repetition period.
Fig.\ref{variance_vs_t}(b) shows, in parallel, the timetrace $V_{W}(t)$
of the normalized variance. During an emission period, 
$V_{W}$ stays below 1, and the statistics of the number of the detected 
photons per pulse is subpoissonian. On the other hand, when the molecule stops 
to emit, the background light yields $V_{W}\approx 1$. 
If we now consider the whole set of data, our measurements yields a single value for the variance 
$V=0.9555$.
In this intensity fluctuation analysis at the level of a single pulse,
this value of $V$ is also directly related to the Mandel 
paramater \cite{Loudon} $Q\equiv\langle(\Delta n)^{2}\rangle/\langle 
n\rangle-1$ by $Q=V-1=-0.0445$.
Let us point out that due to the photodetection deadtime, 
the triggered reference source (R) also yields a subpoissonian counting statistics. 
More precisely, for the coherent source (C) giving about the same mean 
number $\overline{n}_{\rm C}$ of photons per pulse than our SPS, one predicts a value $\displaystyle 
Q=-\overline{n}_{\rm C}/2=-0.0231$. This is confirmed by our measurements on the reference source 
(see Table~\ref{tab_compar}). Nevertheless, the fluctuations of the number of detected photons per 
pulse coming out of our SPS show a clear departure from the reference coherent source. 
Albeit still limited by the quantum efficiency $\eta$, 
this direct measurement of $|Q|$ is larger by more than one order of magnitude compared to previous 
experiments \cite{Short_Mandel_PRL83,Diedrich_PRL87}. For such a solid state SPS like 
ours, any improvement achieved in the light collection efficiency 
would therefore yield higher values of this subpoissonian 
character. We indeed observed, in preliminary experiments, an increase of the collection efficiency
by placing the molecule at a controlled distance of a metallic mirror. 

However, the leak in the dark triplet state induces correlations between consecutive pulses.
The measurement of the variance $V$ of the detected photon number per pulse 
is therefore insufficient to characterize the noise properties of our 
SPS. Whereas such a characterization is usually infered from the record of noise power 
spectra, our photocount measurements are performed in the time domain.
We therefore introduce, as a new variable, the number $N(T)=\sum_{i=1}^k n_i$ 
of the detected 
photons during any period of observation $T=k\tau_{\rm rep}$.
The analysis of the fluctuations of the variable $n_i$ can 
be generalized to the variable $N(T)$, by using the time dependent Mandel parameter \cite{Mandel_OL79} 
$Q(T)\equiv \langle (\Delta N)^2\rangle_{T}/\langle N\rangle_{T} - 1$.
We can also define a Mandel parameter $Q_{\rm s}(T)$ for the number of photons 
\emph{emitted} by the source in the same period of time $T$. 
In the case of an ideal SPS, 
we have $Q=\eta\times Q_{\rm s}$ \cite{Abate_PRA76}.
For such a source, $Q_{\rm s}=-1$, and therefore $Q(T)=-\eta$, for any value of $T$. 
\begin{figure}[h]
\includegraphics[width=8cm]{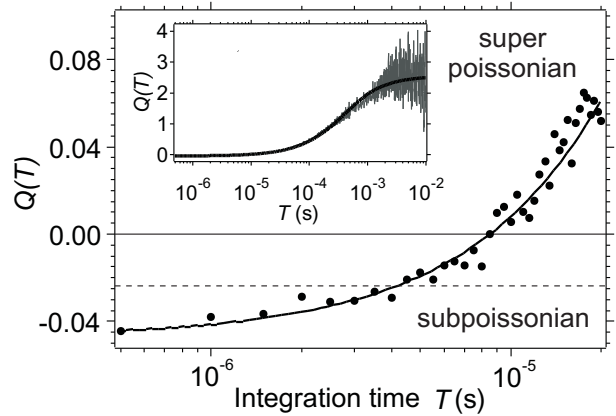} 
\caption{Direct measurements of Mandel parameter $Q(T)$ over short 
recording time $T$. Dashed line shows $Q(T)$ for the equivalent coherent source (C). 
Inset shows $Q(T)$ for longer time of observation. The solid curve is a fit 
to the data by a model accounting for intermittency in the SPS emission.}
\label{Q_vs_T}
\end{figure}
Fig.\ref{Q_vs_T} shows that we did observe subpoissonian intensity fluctuations 
on timescales from $T=1\times \tau_{\rm rep}$ to $T=8\times \tau_{\rm rep}$,
with the minimum value $Q(\tau_{\rm rep})=-0.0445$ achieved on a single pulse 
timescale, as explained above.
When we consider the number of detected photons on timescales larger
than $10^{-5}$~s, the intensity fluctuations 
exhibit a superpoissonnian behaviour ($Q(T)>0$) as shown on inset of Fig.\ref{Q_vs_T}.
This is a direct consequence of the bunching due to the triplet state \cite{Bernard93}.
We developped a simple model to account for the intermittency of the SPS emission. 
In this model, the molecule is either available for fluorescence and 
is said to be in a ON state, or it is in its triplet OFF state and 
does not fluoresce. Let us note $p$, the probability per unit of time to make a 
ON~$\rightarrow$~OFF transition, and $q=1/\tau_{\rm T}$ the one to make the 
reverse OFF~$\rightarrow$~ON transition, where $\tau_{\rm T}$ is the 
lifetime of the triplet state. Note that $p\tau_{\rm rep}=\rm 
\cal{P}_{\rm ISC}$ is the intersystem crossing probability per excitation pulse.
From measured values at the single molecule level with DiIC$_{18}$(3) 
cyanine dye \cite{Veerman_PRL99}, 
$p\tau_{\rm rep}\approx 10^{-4}\ll 1$ and $q\tau_{\rm rep}\approx 
2.5\times 10^{-3}\ll 1$. In this limiting case, the Mandel parameter 
of the source is
\begin{equation}
    Q_{\rm s}(k\tau_{\rm rep})={{2\times\cal{P}_{\rm ISC}}\over \beta^{2}}
    \biggl\{
    1-{1\over{k\beta}}\left[ 1-(1-\beta)^k\right]
    \biggr\} 
    - 1
    \label{model_interm}
\end{equation}
where $\beta\equiv (p+q)\tau_{\rm rep}=\cal{P}_{\rm ISC}+\tau_{\rm rep}/
\tau_{\rm T}$. The Mandel parameter of the detected photon counts is then 
$Q(T)=\eta\times Q_{\rm s}(T)$. As shown on Fig.\ref{Q_vs_T}, our data for $Q(T)$ 
are well fitted by eq.(\ref{model_interm}) over more than 4 orders of magnitude, with
$\eta=0.0445$ (measured) and the free parameters $p$ and $q$.
The fit yields $p\tau_{\rm rep}\approx 2\times 10^{-4}$ and 
$\tau_{\rm T}\approx 250\,\mu$s, in good agreement with Ref.\cite{Veerman_PRL99}.

As a conclusion, the record of every photocount time allows one to 
make a direct time domain fluctuation analysis, as presented in this 
Letter.
This technique can be straightforwardly applied to other SPS, and is also suited 
to investigate photochemical propreties at the single molecule level.
\begin{acknowledgments}
We are grateful to Carl Grossman and Philippe Grangier for help and stimulating discussions.
This work is supported by France Telecom R\&D and ACI ``jeunes 
chercheurs'' (Minist\`ere de La Recherche et de 
l'Enseignement Sup\'erieur).
\end{acknowledgments}


\end{document}